\documentclass[epj,draft]{svjour}
\input{epsf}
\begin{document}
\title{Two interacting particles in a disordered chain I :\\
Multifractality of the interaction matrix elements}
\titlerunning{TIP I: Multifractality of the interaction matrix elements}
\author{Xavier Waintal \and Jean-Louis Pichard}
\institute{CEA, Service de Physique de l'Etat Condens\'e, \\
Centre d'Etudes de Saclay, F-91191 Gif-sur-Yvette, France}
\date{today}
\abstract{For $N$ interacting particles in a one dimensional  
random potential, we study the structure of the 
corresponding network in Hilbert space. The states without 
interaction play the role of the ``sites''. The hopping terms 
are induced by the interaction. When the one body states are 
localized, we numerically find that the set of directly 
connected ``sites'' is multifractal. For the case of two 
interacting particles, the fractal dimension associated to 
the second moment of the hopping term is shown to characterize 
the Golden rule decay of the non interacting states and the 
enhancement factor of the localization length.  
\PACS{{05.45.+b}{Theory and models of chaotic systems}\and 
{72.15.Rn}{Quantum localization}\and
{71.30.+h}{Metal-insulator transitions and other electronic transitions}}
}
\maketitle

  The wave functions of one particle in a random potential have 
been extensively studied. In two dimensions within the localization 
domains~\cite{fe} the large fluctuations of their amplitudes have 
a multifractal character. In one dimension, the elastic mean 
free path $l$ and the localization length $L_1$ coincide, preventing  
a single one particle wave function to be multifractal over a significant 
range of scales. The description of the correlations existing between 
the localized eigenstates is more difficult. This is quite unfortunate, since 
a local two-body interaction re-organizes the non interacting electron 
gas in a way which depends on the spatial overlap of (four) different 
one particle states. When one writes the $N$-body Hamiltonian in the 
basis built out from the one particle states (eigenbasis without 
interaction), this overlap determines the interaction matrix elements, 
i.e. the hopping terms of the corresponding network in Hilbert space. 
In this work, we numerically study the distribution of the hopping 
terms in one dimension, when the one body states are localized. 
It has been observed~\cite{fmpw} that this distribution is 
broad and non Gaussian. We give here numerical evidence that this 
distribution is multifractal. Moreover, since the obtained R\'enyi 
dimensions do not depend on $L_1$, simple power laws describe how 
the moments scale with the characteristic length $L_1$ 
of the one body problem. Since the main applications we consider 
(Golden rule decay of the non interacting states, enhancement factor 
of the localization length for two interacting particles) depend on 
the square of the hopping terms, we are mainly interested by the scaling 
of the second moment. For a size $L \approx L_1$, we show that, contrary 
to previous assumptions, the $N$-body eigenstates without interaction 
directly coupled by the square of the hopping terms have not a density 
of the order of the two-body density $\rho_2(L_1) \propto L_1^{2}$, but 
a smaller density $\rho_2^{\rm eff}(L_1) \propto 
L_1^{f(\alpha(q=2))}$. The dimension $f(\alpha(q=2))$ ($\approx 1.75$ 
for hopping terms involving four different one body states) characterizes 
the fractal set of $N$-body eigenstates without interaction which are 
directly coupled by the square of the hopping terms. 
 
 We consider $N$ electrons described by an Hamiltonian including 
the kinetic energy and a random potential, plus a two-body interaction:
\begin{equation}
H = \sum_{\alpha \sigma} \epsilon_{\alpha} d^+_{\alpha \sigma} 
d_{\alpha \sigma} + U \sum_{\alpha \beta \gamma \delta} 
Q_{\alpha \beta}^{ \gamma \delta}  d^+_{\alpha \downarrow}  
d_ {\gamma \downarrow}  d^+_{\beta \uparrow}  d_{\delta \uparrow}
\end{equation}
 The operators $d^+_{\alpha \sigma}$ ($d_{\alpha\sigma}$) create (destroy) 
an electron in a one body eigenstate $|\alpha>$ of spin $\sigma$. Noting 
$\Psi_{\alpha}(n)$ the amplitude  on site $n$ of the state $|\alpha>$ 
with energy $\epsilon_{\alpha}$, the interaction matrix elements 
are proportional to the $Q_{\alpha \beta}^{ \gamma \delta}$ given by:

\begin{equation}\label{Q}
Q_{\alpha \beta}^{ \gamma \delta}= 
\sum_n \Psi^*_{\alpha}(n) \Psi^*_{\beta}(n) \Psi_{\gamma}(n) \Psi_{\delta}(n)
\end{equation}
 
 This comes from the assumption that the interaction 
$U \sum_n c^+_{n\downarrow} c_{n\downarrow} c^+_{n\uparrow} 
c_{n\uparrow}$ is local. The $c^+_{n\uparrow}$ 
($c_{n\sigma}$) create (destroy) an electron on the site $n$ and  
$ d^+_{\alpha \sigma} = \sum_{n} \Psi_{\alpha}(n) c^+_{n \sigma} $. 
When $U=0$, the Hamiltonian is diagonal in the basis built out from 
the one particle states, and the $N$ body states ($\prod_{i=1}^{N} 
d^+_{\alpha_i \sigma_i}|0>$) can be thought as the ``sites'' with energy 
$\sum_{i=1}^{N} \epsilon_{\alpha_i}$ of a certain network which is 
not defined in the real space, but in the $N$-body Hilbert space. 
When $U \neq 0$, different ``sites'' can be directly connected by 
off-diagonal interaction matrix elements. Therefore, one can 
map~\cite{agkl} this complex $N$-body problem onto an Anderson 
localization problem defined on a particular network in the $N$-body  
Hilbert space. Since the interaction is two body, only 
the ``sites'' differing by two quantum numbers can be directly coupled. 
This restriction will not matter~\cite{sh} for $N=2$ and 
(under certain approximations) may yield a Cayley tree topology~\cite{agkl} 
for the resulting network, if $N$ is large. We study the 
additional restrictions coming from one body dynamics. 

 We summarize a few evaluations of the second moment ($q=2$) of 
$Q_{\alpha\beta}^{\gamma\delta}$ which have been previously used. 
Case (i): The one body Hamiltonian is described by random matrix 
theory (RMT). The statistical invariance under orthogonal transformations 
$O(M)$ implies that $\langle(Q_{\alpha\beta}^{\gamma\delta})^2\rangle \approx 
1/M^{3}$ where $M$ is the number of one body states. Case (ii): The 
system is a disordered conductor of conductance $g$. An estimate~\cite{agkl} 
based on perturbation theory gives $\langle 
(U Q_{\alpha\beta}^{\gamma\delta})^2 \rangle \propto (\Delta / g )^2$. 
Since the one particle mean level spacing $\Delta \propto 1/M$, this 
perturbative result coincides with the previous RMT results if one takes 
$M \equiv g^2$. Moreover, it is valid only if all the one particle 
states appearing in Eq.(\ref{Q}) are taken from a 
sequence of $g$ consecutive levels in energy. Otherwise, 
$Q_{\alpha\beta}^{\gamma\delta}$ can be neglected. Case (iii): 
The system is a disordered insulator. Shepelyansky~\cite{sh} in his first 
study of the two interacting particles (TIP), assumes a 
RMT behavior for the $M=L_1^d$ components of the wave function 
inside the localization domain, and neglects the exponentially 
small components outside this domain. When the dimension $d=1$, one gets 
a term $(Q_{\alpha\beta}^{\gamma\delta})^2 \approx 1/L_1^{3}$ for 
the terms coupling a TIP state $|\alpha\beta>$ to $L_1^2$ TIP states 
$|\gamma\delta>$. This estimate for $g<1$ differs from the one valid 
when $g>1$ under two important aspects: not only $M \approx L_1^d$ instead 
of $g^2$, but the condition for a large hopping is entirely different. 
In the insulator, a large hopping term is not given by four one particle 
states close in energy, but by four states close in real space, i.e. 
located inside the same localization domain. Ponomarev and Silvestrov 
have criticized~\cite{ps} this estimate, using an approximate description 
of a localized state for weak disorder. They note that the density of TIP 
states coupled by the interaction is sensibly smaller.   

 For a more accurate study of $Q_{\alpha \beta }^{\gamma \delta}$ in  
one dimension, we consider a spin independent one particle Anderson 
tight binding model with $L$ sites and nearest neighbor hopping 
($t \equiv 1$). The on-site potentials $V_n$ are taken at 
random in the interval [-W,W] and the boundary conditions are periodic. 
$L_1$ is estimated from the weak disorder formula 
$L_1\approx 25/W^2$ . The $|Q_{\alpha \beta }^{\gamma \delta}|$ are 
calculated using Eq.(\ref{Q}) and numerical diagonalization of the one 
particle Hamiltonian. $Q_{\alpha\beta}^{\gamma\delta}$, for fixed $\alpha$ 
and $\beta$ is a two-dimensional object which is not defined in the 
real $2d$ space, but in the space of two one particle quantum numbers 
$\gamma$ and $\delta$. Those states $|\gamma>$ (and $|\delta>$) 
can be ordered in different ways: (a) spectral ordering by increasing 
eigenenergy, (b) spatial ordering by the location $n_{\gamma}$ of 
their maximum amplitude, from one side of the sample to the other,
(c) momentum ordering if $W=0$. Let us note that ordering (b) becomes 
meaningful only in the localized regime ($L>L_1$).

\begin{figure}[tbh]
\epsfxsize=3.5in
\epsfysize=3in
\epsffile{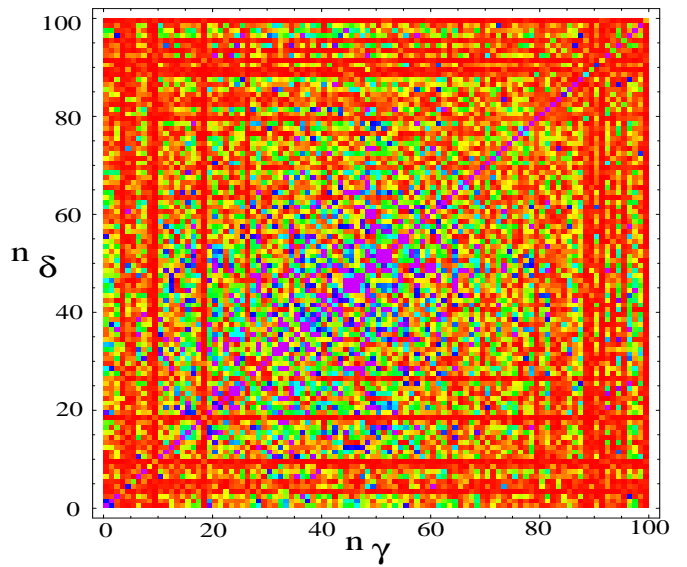}
\vspace{3mm}
\caption[fig1]{\label{fig1}
$|Q_{\alpha_0 \alpha_0}^{\gamma \delta}|$  with $|\alpha_0>$ taken  
in the bulk ($n_{\alpha_0}=50$) of a sample of size $L=L_1=100$. 
Spatial ordering (b). The color code goes from red (small value) 
to violet (large value) through yellow, green and blue.}
\end{figure}

 We first study the matrix element $Q_{\alpha_0\alpha_0}^{\gamma\delta}$, 
characterizing two electrons with opposite spins in the same state 
$|\alpha_0>$ hopping to an arbitrary state $|\gamma\delta>$. 
Hopping is very unlikely over scales larger than $L_1$. 
The $L_1^2$ large values of the hopping term are concentrated 
inside a square of size $L_1^2$, as shown in Fig.\ref{fig1} 
for a given sample using ordering (b) and a rainbow color code. 
Fig.\ref{fig1} is not homogeneously colored, but exhibits a complex 
pattern which reminds us another bi-dimensional object: the one particle 
wave function in a two dimensional disordered lattice. This suggests us 
to analyze its fluctuations as for the $2d$ one body states, and to check if 
this pattern is not the signature of a multifractal structure.

\begin{figure}[tbh]
\epsfxsize=3in
\epsfysize=3in
\epsffile{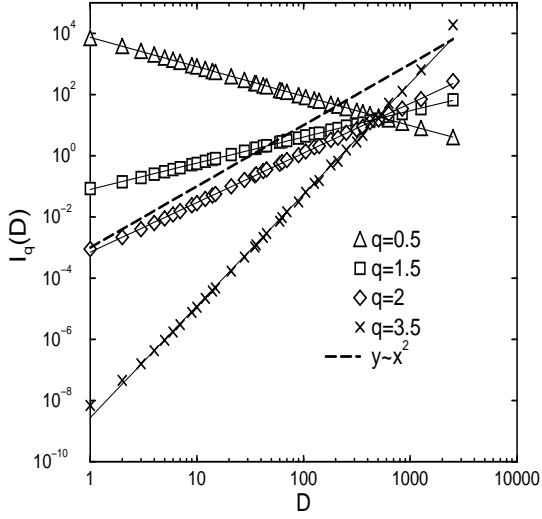}
\vspace{3mm}
\caption[fig2]{\label{fig2a}
Power laws showing that Fig.~\ref{fig1} corresponds to a multifractal 
measure in the $(\gamma,\delta)$ two dimensional plane. $I_q(D)$ 
are calculated for a single sample with $L_1=L=2500$. $\alpha_0$ 
in the bulk of the spectrum. The states $|\gamma\rangle$ and $\delta\rangle$ 
are ordered by increasing eigenergy (ordering a). The dashed line 
corresponds to the RMT prediction (case (i)). }
\end{figure}
 In analogy with the $2d$ one body problem, we do not expect that this 
multifractality will be valid in the whole $(\gamma,\delta)$ Hilbert space, 
but only in a limited but parametrically large domain. 

 We proceed as usual (see references \cite{pv,hjkps}) for the 
multifractal analysis. For $L_1$ and $L$ fixed, we divide the 
plane $(\gamma,\delta)$ into $(L/D)^2$ boxes of size $D$ and 
we calculate the ensemble averaged function for different values of $q$
\begin{equation}\label{I}
  I_q(D) = \sum_{i=1}^{N_{boxes}} 
\left( \sum_{\gamma,\delta \in \mbox{box}_i} 
| Q_{\alpha_0 \alpha_0}^{\gamma \delta} | \right)^q.
\end{equation}
 The existence of a multifractal measure defined in the 
$(\gamma,\delta)$-plane by the interaction matrix elements 
is established in the next figures. In Fig.\ref{fig2a}, a 
single sample has been used and power laws 
$I_q(D) \propto D^{\tau(q)}$ are obtained over many orders of magnitude  
for different values of q. 

\begin{figure}[tbh]
\epsfxsize=3in
\epsfysize=3in
\epsffile{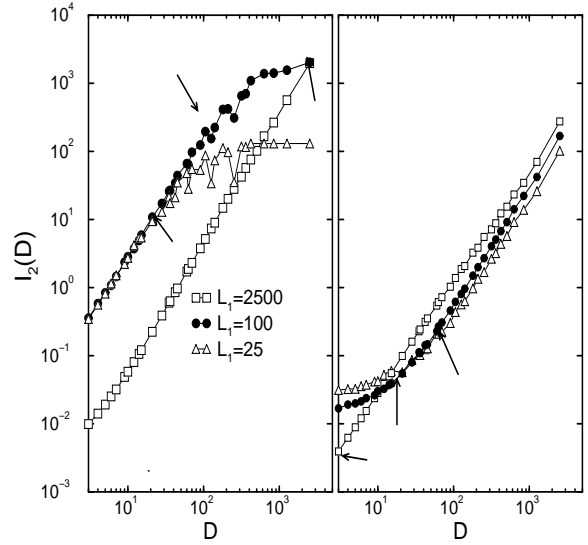}
\vspace{3mm}
\caption[fig2]{\label{fig2b}
Left: Spatial ordering (b). $I_2(D)$ for 
$|Q_{\alpha_0\alpha_0}^{\gamma\delta}|$ with $\alpha_0$ in the bulk 
of the spectrum and for $L_1$ (indicated by 
the arrows) $=25, 200, 2500 \leq L = 2500$. The power law behaviors 
are obtained for a given sample if $1 \leq D \leq L_1$. Right: Same 
samples using spectral ordering (a). The arrows indicates the lower 
scales associated to $\Delta(L_1)$. The power laws are valid for 
$L/L_1 \leq D \leq L$. }
\end{figure}

The limits of validity of these power laws are shown in Fig.~\ref{fig2b}. 

On the left side, spatial ordering (b) 
for different values of $L_1$ is used for the states ($\gamma,\delta$). 
One can see that $I_q(D) \propto D^{\tau(q)}$, for scales $1 < D < L_1$, 
as indicated by the arrows. The lower scale is given by the lattice 
spacing of the $(n_{\gamma},n_{\delta})$ network in Hilbert space.
The upper scale $L_1$ is the largest scale compatible with a spatial 
overlap of the states $\gamma$ and $\delta$, for a fixed $\alpha_0$.
This means that the multifractality of the interaction matrix elements 
$Q_{\alpha_0\alpha_0}^{\gamma\delta}$ in the two dimensional Hilbert 
space $(\gamma,\delta)$ has the same parametrically large range of 
validity as the one body wave function~\cite{fe} in two dimensions 
(scale $ 1< D < L_1$). Here, multifractality is valid for $L_1^2$ 
matrix elements as multifractality is valid in the $2d$ one body 
problem for $L_1^2$ sites. 

 On the right side of Fig.~\ref{fig2b} spectral ordering (a) is used 
for the same samples, giving the same power laws as with ordering (b), 
inside the corresponding energy range ($\Delta(L_1) < D \Delta (L) < 1$) 
indicated by the arrows. $\Delta(x) \propto x^{-1}$ is the level spacing of a 
segment of size $x$, and 1 is the band width. The exponents $\tau(q)$ are 
independent of the ordering when $L > L_1$ (i.e. when the ordering (b) 
becomes meaningful) and the small fluctuations from sample to sample 
are removed by ensemble averaging. 

\begin{figure}[tbh]
\epsfxsize=3in
\epsfysize=3in
\epsffile{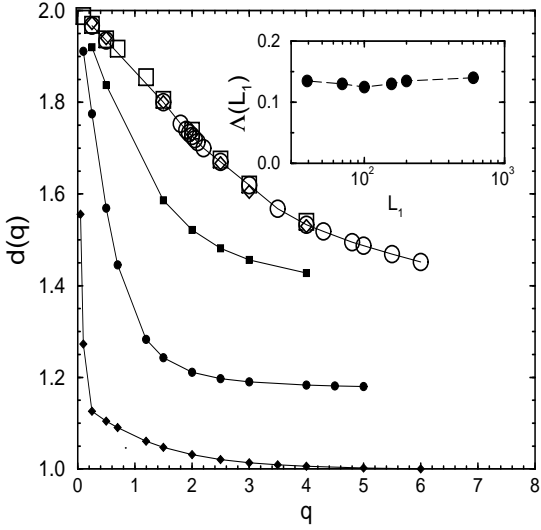}
\vspace{3mm}
\caption[fig2]{\label{fig2c}
 $d(q)$ with $\alpha_0$ in the bulk of the spectrum using 
ordering (a) and after ensemble averaging. Filled symbols: $L=240$ and 
$L_1=\infty$ (diamonds), $25.10^4$ (circles) and $2500$ (squares). 
Open symbols: $L_1 = 70$ and $L=960$  (diamonds), $480$ (circles) and 
$240$ (squares). Insert:  the slope $\Lambda (L_1)$ showing that $d(q)$ is 
disorder independent for $q \leq 3$ and $L_1 < L$.}
\end{figure}

The corresponding R\'enyi dimensions 
$$
d(q)\equiv \tau(q)/(q-1)
$$ 
are shown in Fig.~\ref{fig2c} for different $L$ and $L_1$, using ordering 
(a) and ensemble averaging. 

 For an infinite $L_1$ (no disorder), the eigenstates are plane waves of 
momentum $k_{\alpha}$ and $Q_{\alpha \beta}^{ \gamma \delta} \neq 0$ 
only if $k_\alpha + k_\beta - k_\gamma - k_\delta=0$. This gives 
$d(0)=2$ and $d(q>0)=1$ with ordering (c). The dimensions calculated 
with ordering (a) are close to this limit. 
For a finite $L_1$, $d(q)$ goes from the clean limit ($L << L_1$) 
to an $L_1$-independent regime when $L >> L_1$. In the crossover regime 
($L \leq L_1$) the $d(q)$ depend on $L_1$. In the limit $L >> L_1$, 
the $d(q)$ (using orderings (a) or (b)) do not depend on $L$ and $L_1$. 
For $0 < q \leq 3$, 
$$
d(q) \approx 2-\Lambda q
$$ 
with a slope $\Lambda \approx 0.135$. The $L_1$-independence of $\Lambda$ 
is shown in the insert of Fig.~\ref{fig2c} for $L_1\leq L$ up to $L_1 = 600$. 

\begin{figure}[tbh]
\epsfxsize=3in
\epsfysize=3in
\epsffile{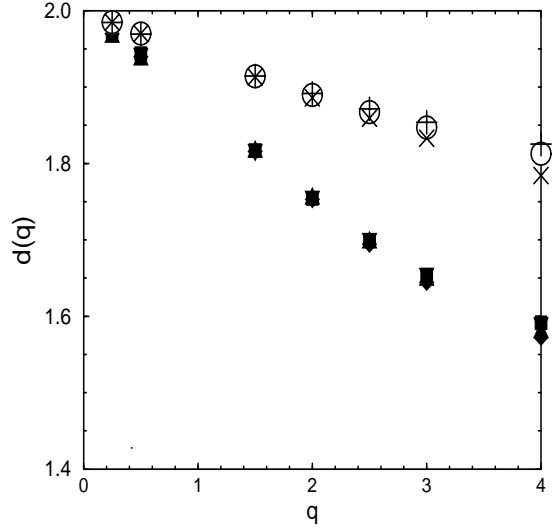}
\vspace{3mm}
\caption[fig2]{\label{fig2d}
d(q) calculated for $L_1=70$ and $L=240$ using ordering (a). 
The filled symbols  correspond to $\alpha=\beta=\alpha_0$, 
averaged over 10 consecutive $\alpha_0$ chosen in various parts 
of the spectrum: down triangle ($130 \leq \alpha_0 \leq 140$), 
square ($120 \leq \alpha_0 \leq 130$), diamond ($145 \leq \alpha_0 
\leq 155$) and up triangle ($170 \leq \alpha_0 \leq 180$). 
Empty symbols correspond to the case $\alpha \neq \beta$. 
$\alpha$ is fixed and $d(q)$ is averaged for a few $\beta \neq 
\alpha$: circle ($\alpha=100, 123 \leq \beta \leq 133$), 
cross  ($\alpha=80, 123 \leq \beta \leq 133$) and 
plus  ($101 \leq \alpha \leq 112$ and $123 \leq \beta \leq 133$)}
\end{figure}
  A multifractal distribution has scaling behavior described by 
the $f(\alpha)$-spectrum, given by the relations:

\begin{equation}{\label{LT}}
\alpha(q)=\frac{d\tau}{dq} {\rm \ \ 
and\   \ }f(\alpha(q))=\alpha(q).q-\tau(q).
\end{equation} 

We obtain 
\begin{equation}
f(\alpha(q))\approx 2-\Lambda q^2
\end{equation}
for $q \leq 3$, i.e. a parabolic shape 
$f(\alpha)=2-(\alpha-2-\Lambda)^2 / (4 \Lambda)$ 
around the maximum $2+\Lambda$. We have mainly studied the first positive 
moments, since we are mainly interested by $f(\alpha(q=2))$. Indeed, 
when one uses Fermi golden rule to calculate the interaction-induced 
decay of a non-interacting state, one needs to know the density of 
states directly coupled by the second moment $(q=2)$ of the hopping 
term. The fractal dimension of the support of this density is given 
by $f(\alpha(q=2))$. For greater values of $q$, there are deviations 
around the parabolic approximation, indicating deviations 
around simple lognormal distributions. From a study of the large and 
small values of $|Q_{\alpha_0\alpha_0}^{\gamma\delta}|$, one can 
obtain $ d(q\rightarrow\pm\infty) $. We find $d(+\infty)=1.33$ 
and $d(-\infty)=3.15$, giving the limits of the support of $f(\alpha)$.

 We have also checked that our results for 
$Q_{\alpha_0\alpha_0}^{\gamma\delta}$ do not depend on the chosen $\alpha_0$ 
and studied the general case where $|\alpha>$ and $|\beta>$ 
are not the same. In Fig.~\ref{fig2d}, one can see that the 
$Q_{\alpha_0\alpha_0}^{\gamma\delta}$ studied for different $\alpha_0$ 
give the same curves $d(q)$. 
Using energy ordering (a) and imposing an energy separation 
$|\epsilon_{\alpha} - \epsilon_{\beta}| > \Delta(L_1)$ in order to 
have a good overlap between the fixed states $\alpha$ and $\beta$, 
we find also power law behaviors for $I_q(D)$. The 
corresponding dimensions $d(q)$ are given in Fig.~\ref{fig2d}, 
characterized by a slope 
$$
\Lambda (\alpha \neq \beta) \approx \Lambda(\alpha=\beta)/2 \approx 0.065.
$$ 
Therefore, the multifractal character of $ Q_{\alpha\beta}^{\gamma\delta}$ 
is less pronounced when $|\alpha> \neq |\beta>$, but remains relevant.

 So far, we have discussed the hopping terms of the general $N$-body 
problem. We now discuss how our results modify previous assumptions 
for two interacting particles (TIP). As pointed out by Shepelyansky, 
the interaction induced hopping mixes nearby in energy TIP states 
$|\alpha\beta>=d^+_{\alpha\downarrow} d^+_{\beta\uparrow}|0>$. The 
decay width $\Gamma$ \cite{js,wp,wpi} of a TIP state $|\alpha\beta>$, 
built out from two one particle states localized within $L_1$, can be 
estimated using Fermi golden rule.  If one assumes RMT wave functions 
inside $L_1$ for the one particle states, (case (iii)) the  
$ Q_{\alpha \beta }^{\gamma \delta} \approx \pm U.L_1^{-3/2}$ couple 
the TIP state $|\alpha\beta>$ to all the TIP states $|\gamma\delta>$ 
inside $L_1$. Around the band center, they have a density $\rho_2(L_1) 
\propto L_1^2$ and  Fermi golden rule gives 
\begin{equation}
\Gamma (E \approx 0) \propto \frac{U^2}{L_1^3} \rho_2(L_1) = \frac{U^2}{L_1}
\end{equation}. 
We have shown that all 
the TIP states which can be coupled by the interaction within the 
localization domains are not equally coupled. Since the square 
of the hopping terms appears in the Golden rule, our multifractal 
analysis gives a reduced effective TIP density $\rho_2^{\rm eff} 
\propto L_1^{f(\alpha(q=2))}$ which should replace the total 
TIP density $\rho_2(L_1)$. The resulting expression
\begin{equation}{\label{gamma}}
\Gamma_{\alpha \beta}\propto \frac{U^2}
{L_1^{3}} L_1^{f(\alpha(q=2))} 
\end{equation}
can be compared to the direct numerical evaluation: 
\begin{equation}{\label{GR}}
 \Gamma_{\alpha \beta} = U^2\sum_{\gamma \delta} 
|Q_{\alpha \beta}^{ \gamma \delta}|^2 \delta(\epsilon_{\alpha} 
+ \epsilon_{\beta} - \epsilon_{\gamma} - \epsilon_{\delta})
\end{equation}
of the Golden rule decay. 
 
\begin{figure}[tbh]
\epsfxsize=3in
\epsfysize=2.5in 
\epsffile{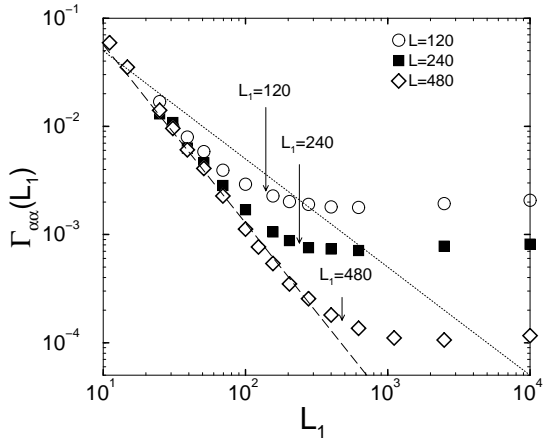}
\vspace{3mm}
\caption[fig3]{\label{fig3}
$\Gamma_{\alpha\alpha}(L_1)$ for three different sizes $L$. Dashed
line: $y=2.55 L_1^{-1.65}$. Doted line: $y=0.5 L_1^{-1}$}
\end{figure}

In Fig.\ref{fig3}, we show for three 
different sizes $L$ how the decay rate $ \Gamma_{\alpha \alpha} $ 
numerically calculated using Eq.(\ref{GR}) depends on $L_1$, 
for a TIP state $|\alpha\alpha>$ where $\alpha$ is taken in the 
bulk of the spectrum. From Fig.\ref{fig2c}, one gets $\alpha(2)\approx 
1.52$ and $\tau(2)\approx 1.69$, and Eq.(\ref{LT}) gives 
$f(\alpha(q=2)) \approx 1.35$. For this value, one can see in 
Fig.\ref{fig3} that Eq.(\ref{gamma}) and Eq.(\ref{GR}) give indeed 
the same $L_1$-dependence. This observed $L_1^{-1.65}$ law clearly 
differs from the $L_1^{-1}$ law implied by the RMT assumption 
(case (iii)). We can also see that $\Gamma_{\alpha\alpha}$ does not 
depend on $L$ when $ L_1 < L$, since there are no significant 
hopping terms for range larger than $L_1$. 

 Another interesting issue is the enhancement of the localization 
length $L_2$, which is induced by the interaction and characterizes 
a restricted set of TIP states which have a sufficient overlap to 
be re-organized by a local interaction \cite{owm,wmpf}. Using the 
Thouless block scaling analysis~\cite{i}, one finds $ \frac{L_2}{L_1} 
\propto (\pm U /L_1^{3/2} \rho_2 (L_1))^2 $. If the density $\rho_2(L_1)$ 
of states coupled by the interaction is the total TIP density for a 
size $L_1$, one finds the original estimate~\cite{sh,i} 
$L_2 \propto L_1^2$. The multifractality yields a reduced 
effective density $\rho_2^{\rm eff} \propto L_1^{f(\alpha(q=2))}$ 
instead of the total TIP density. Since the contribution 
of TIP states $|\alpha\beta>$ with $\alpha \neq \beta$ dominates, 
we use $f(\alpha(2))=1.75$ valid when $\alpha \neq \beta$ and we 
find $L_2 \propto L_1^{1.5}$. This $L_1$-dependence is in agreement 
with recent numerical results~\cite{fmpw,sk}. So there is an enhancement, 
though weaker than the original estimate~\cite{sh} ($L_2 \propto L_1^2$), 
due to the multifractal distribution of the hopping terms. 

  In summary, we have studied how one particle dynamics 
(one dimensional localization) can affect the many body 
problem through non trivial properties of the distribution 
of the two-body interaction. In a clean system, one has 
$f(\alpha(2))=1$ and the density of states which are effectively 
coupled by the interaction is the one particle density $\rho_1 \propto 
L$. The disorder, as it is well known, enhances the effect of the 
interaction, since the effective density $\rho_2^{\rm eff} 
\propto L_1^{f(\alpha(2))}$, with $1<f(\alpha(q=2)<2$ for $L=L_1$. 
This enhancement of the density of states coupled by the interaction 
inside a system of size $\approx L_1$ is nevertheless smaller than the one 
($\rho_2\propto L_1^2$) given by fully chaotic one body states inside 
their localization domains. In a second paper~\cite{wwp}, a study of the 
TIP spectral fluctuations will be presented, showing that statistics 
is critical (as for the one body spectrum at a mobility edge) if $U$ is 
large enough, accompanied by multifractal wavefunctions in the TIP 
eigenbasis for $U=0$. In a third paper~\cite{dtawp}, a study of the 
dynamics of a TIP wave packet will be presented, showing that the center of 
mass exhibits anomalous diffusion between $L_1$ and $L_2$. These three 
studies provide consistent and complementary observations supporting our 
claim: multifractality and criticality are 
relevant concepts for a TIP system with on site interaction in one dimension. 
Our results go beyond the TIP problem and show that 
oversimplified two-body random interaction matrix models~\cite{bf,fic,js2} 
which ignore multifractality in the hopping cannot properly describe the 
many body quantum motion in Anderson insulators.
 
 We are indebted to S. N. Evangelou for very useful comments.

\end{document}